\documentclass[12pt]{iopart}
\usepackage[latin1]{inputenc}
\usepackage[T1]{fontenc}
\usepackage[english]{babel}
\usepackage{latexsym}
\begin{document}
\title[Sufficient conditions for curvature invariants to avoid divergencies in...]{Sufficient conditions for curvature invariants to avoid divergencies in Hyperextended Scalar Tensor theory for Bianchi models}
\author{Stéphane Fay\footnote{Email: Steph.Fay@Wanadoo.fr}}
\address{14 rue de l'\'Etoile\\
75017 Paris\\
France}
\begin{abstract}
We look for sufficient conditions such that the scalar curvature, Ricci and Kretchmann scalars be bounded in Hyperextended Scalar Tensor theory for Bianchi models. We find classes of gravitation functions and Brans-Dicke coupling functions  such that the theories thus defined avoid the singularity. We compare our results with these found by Rama in the framework of the Generalised Scalar Tensor theory for the FLRW models.
\\
\\
\\
Published in Classical and Quantum Gravity copyright 2000 IOP Publishing Ltd\\
Classical and Quantum Gravity, Vol 17, 14, 2000.
\end{abstract}

\maketitle
\section{Introduction} \label{s1}
An important problem in cosmology is the presence of singularities, i.e. a set of points in spacetime where the laws of classical physics would be broken. This problem has been studied in numerous papers. One of the most famous is this of Hawking and Penrose \cite{HawEll73}. It is shown that, for the FLRW models with matter field respecting the strong and weak energy conditions, it always exists a singularity. Some methods have also been developed to build Lagrangien such that the theory hence defined be non singular \cite{Bra93,BraMukSor93,Bra95}. Another method is used in quantum cosmology where the absence of singularity is sometimes imposed by writing that the wave function vanishes with the scale factor. Last, for the string theory, Gasperini and Veneziano have proposed the models of Pre-Big-Bangs which could allow to avoid the singularity \cite{Gas94,Gas99}.

In any case, to get a non singular theory it is necessary that all the curvature invariants be bounded. This point has been studied by Rama in  \cite{Ram97} for the FLRW models and the Generalised scalar tensor theory (GST). In this paper we wish to examine from the same viewpoint what is the situation in the Hyperextended scalar tensor theory (HST) for the Bianchi models by studying the divergence of the scalar curvature, Ricci and Kretchmann scalars which are the most common curvature invariants met in the literature. Note that this type of study is made at a classical level whereas singularity deals with quantum cosmology. However, we hope that the absence of singularity at a classical level would indicate their absence at a quantum one.

Let us justify the geometrical framework of this paper. Although for present time our Universe seems to be isotropic, it is not proved that it was the case at early times or even that it is not a local phenomenon. Then, it is interesting to consider the homogenous models, i.e. the Bianchi models. Among them, the Bianchi types $I$, $V$, $VII_0$, $VII_h$ and $IX$ models, which admit FLRW solutions, are able to isotropize \cite{ChaCer95}. Hence we will study the Bianchi type $I$ and $V$ models.  When they isotropize, the first one tends toward the flat isotropic model and the second one toward the open one. We will also study the Bianchi type $VI_0$ model, considered in \cite{ChiLab98,BarGasSaf99}. Last we will examined the Bianchi type $II$ model which is representative of the Bianchi models of class A during phases of strong anisotropy \cite{Lid96}.

Let us justify the study of the HST \cite{TorVuc96,Tor97}. Its Lagrangian contains two free functions depending on a scalar field $\phi$. The first one, $G(\phi)$, represents the gravitational function and the second one, $\omega(\phi)$, a coupling function between the scalar field and the metric. The scalar fields are predicted by particle physics theories as string theory or supergravity. In cosmology they allow to solve numerous difficulties as age problem or inflationary exit. However, the use of theories with free functions depending on $\phi$ is also the source of new problems: what are the classes of functions $G$ and $\omega$ which are agreed with both observational tests \cite{SerAli96,SanKalWag97,Tor99} and theoretical considerations such as the absence of singularity. In this work we will consider this last question: our goal is to find sufficient conditions on $G$ and $\omega$ such that some curvature invariants do not diverge. We will not consider other forms of  matter but scalar fields since scalar field dominated models are often asymptotical solutions for early or late times.

The paper is organised as follows. In section \ref{s2}, we write the field equations of the HST, the scalar curvature, Ricci and Kretchmann scalars. In section \ref{s3}, we determine sufficient conditions such that they be bounded at any times for the Bianchi type $I$, $II$, $V$ and $VI_0$ models. In section \ref{s4}, we use them to determine some suitable forms of $\omega$ for the GST and a string inspired theory. We conclude in section \ref{s5} and compare our results with these of Rama.
\section{The curvature invariants} \label{s2}
We use the following line element:
\begin{equation} \label{1}
ds^{2}=-dt^{2}+e^{2\alpha} (\omega^1)^2+e^{2\beta}(\omega^2)^2+e^{2\gamma}(\omega^3)^2
\end{equation}
The $\omega^i$ are the one forms specifying each Bianchi model. The Lagrangien of the HST is written:
\begin{equation} \label{2}
L=G(\phi)^{-1}R-\frac{\omega(\phi)}{\phi}\phi_{,\mu}\phi^{,\mu}
\end{equation}
where $\phi$ is the scalar field, $\omega$ the coupling function and $G$ the gravitation function, both depending on $\phi$. We get the field equations and the Klein-Gordon equation by varying the action with respect to the metric functions and the scalar field:
\begin{eqnarray} \label{3}
R_{\mu\nu}-\frac{1}{2}g_{\mu\nu}R & = & G\mbox{[}\frac{\omega}{\phi}\phi_{,\mu}\phi_{,\nu}-\frac{\omega}{2\phi}\phi_{,\lambda}\phi^{\lambda}g_{\mu\nu}+(G^{-1})_{,\mu;\nu}-g_{\mu\nu}\Box (G^{-1})\mbox{]}
\end{eqnarray}
\begin{equation} \label{4}
\dot{\phi}^{2}\left[-\frac{\omega_{\phi}}{\phi}+\frac{\omega}{\phi^{2}}-G(G^{-1})_{\phi}\frac{\omega}{\phi}\right]+\frac{2\omega}{\phi}\Box \phi+3G(G^{-1})_{\phi}\Box G^{-1}=0
\end{equation}
An overdot means a derivative with respect to the proper time $t$. To calculate the curvature invariants, we define the $\tau$ time by $dt=e^{\alpha+\beta+\gamma}d\tau$. It would be more interesting to use the proper time $t$, since $\tau$ is not a physically significant times. However calculus in the Bianchi model are more tractable in the $\tau$ time. In fact, we will first get our results in the $\tau$ time and will generalise then in the $t$ times in the last section by making comparisons with the results of Rama for the FLRW models.\\
\\
The first curvature invariant we compute is the scalar curvature, obtained by contracting the equation (\ref{3}):
\begin{equation} \label{6}
R=V^{-2}G(-\omega\phi^{-1}\phi'^{2}-3(G^{-1})'')
\end{equation}
The prime holds for derivative with respect to $\tau$ and $V=e^{\alpha+\beta+\gamma}$ defines the 3-volume of the Universe. We introduce (\ref{6}) in (\ref{3}) to obtain an expression for $R_{\mu\nu}$ and then we get the Ricci scalar:
\begin{eqnarray} \label{7}
R_{\mu\nu}R^{\mu\nu} & = & V^{-4}G^2\mbox{[}\omega^2\phi^{-2}\phi'^{4}+\omega\phi^{-1}\phi'^{2}(3(G^{-1})''-2(G^{-1})'V'V^{-1})\nonumber\\
 & & +(-(G^{-1})''^2+(G^{-1})'^2V^{-2}V'^2-2(G^{-1})''(G^{-1})'V^{-1}V')\mbox{]}
\end{eqnarray}
Last, the Kretchmann scalar defined as $R^{\alpha\beta\mu\nu}R_{\alpha\beta\mu\nu}$ will be calculated with the help of:
\begin{equation}
R_{\alpha\beta\mu\nu}=\Gamma_{\alpha\beta\nu,\mu}-\Gamma_{\alpha\beta\mu,\nu}+\Gamma^{m}_{\beta\nu}\Gamma_{\alpha m \mu}-\Gamma^{m}_{\beta\mu}\Gamma_{\alpha m \nu}-C^{m}_{\mu\nu}\Gamma_{\alpha\beta m}
\end{equation}
with
\begin{equation}
\Gamma_{\alpha\beta\mu}=1/2(g_{\alpha\beta,\mu}+g_{\alpha\mu,\beta}-g_{\beta\mu,\alpha}+C_{\mu\alpha\beta}+C_{\beta\alpha\mu}-C_{\alpha\beta\gamma})
\end{equation}
The $\Gamma$ are the connections and the $C$ the structure constants specifying each Bianchi model. To express this scalar as a function of $V$, $\omega$, $G$ and $\phi$, as the two previous ones, we need to solve the field equations (\ref{3})-(\ref{4}) to get $\alpha$, $\beta$ and $\gamma$ depending on this quantities. In the next section we choose sufficient conditions such that the three curvature invariants be bounded.
\section{Sufficient conditions such that the scalar curvature, Ricci and Kretchmann scalars be bounded} \label{s3}
The Klein-Gordon equation can be integrated to give:
\begin{equation} \label{5}
\left[\frac{3}{4}(G^{-1})_\phi^{\mbox{ }2}+\frac{1}{2\phi}G^{-1}\omega\right]\phi'^2=\phi_0
\end{equation}
$\phi_0$ is an integration constant. The scalar field is thus a monotonous function of time. The term in square bracket is proportional to the energy density of the scalar field in the Einstein frame. If we assume a positive energy density, we get a variation interval for $\phi$. Moreover, equation (\ref{5}) allows us to write $\phi'$, $(G^{-1})'=G^{-1}_\phi \phi'$ and $(G^{-1})''=((G^{-1})')_\phi \phi'$ as functions of $\omega$ and $G$:
\begin{equation} \label{9}
(G^{-1})'=\phi_0^{1/2}(G^{-1})_\phi (\frac{3}{4}(G^{-1}_{\mbox{   }\phi})^2+\frac{G^{-1}\omega}{2\phi})^{1/2}
\end{equation}
\begin{equation} \label{10}
(G^{-1})''=4\phi_0\frac{2(G^{-1})_{\phi\phi}\omega G^{-1}\phi-(G^{-1})_\phi^{\mbox{   }2}\omega\phi+(G^{-1})_\phi(\omega G^{-1}-G^{-1}\omega_\phi \phi)}{(2G^{-1}\omega+3\phi(G^{-1})_\phi^{\mbox{  }2})^2}
\end{equation}
Our aim being to  choose sufficient conditions on $G$ and $\omega$ such that the curvature invariants do not diverge, we have just to write them as function of $G$, $\omega$, $\phi$ and their derivatives with respect to $\tau$ and to use the expressions (\ref{9}) and (\ref{10}) to achieve our goal. In the first subsection, we look for these sufficient conditions. Sometimes, their expressions depend on the Bianchi type. They will be studied in the second subsection.
\subsection{Sufficient conditions such that the curvature invariants be bounded} \label{s31}
Each of the three curvature invariants depends on the 3-volume. So the first sufficient condition we will choose will be $V\not = 0$. Its expression as a function of the scalar field depends on the Bianchi model and will be studied in the next subsection. Assuming that $V\not=0$, sufficient conditions such that the scalar curvature (\ref{6}) be bounded whatever the Bianchi type will be that the following quantities do not diverge:
\begin{itemize}
\item $G(G^{-1})''$
\item $G\omega\phi'^2\phi^{-1}$
\end{itemize}
Each of them may be expressed as a function of $G$ and $\omega$ independently of the Bianchi type.
\\
For the Ricci scalar, it is sufficient that the following quantities do not diverge:
\begin{itemize}
\item $G(G^{-1})''$
\item $G\omega\phi'^2\phi^{-1}$
\item $G(G^{-1})'$
\item $V'V^{-1}$ i.e. $\alpha'+\beta'+\gamma'$
\end{itemize}
The expression of the last one as a function of $G$ and $\omega$ depends on the Bianchi model and will be studied in the next subsection. The two first conditions have already been chosen for the scalar curvature. The third one is new. Its expression as a function of $G$ and $\omega$ does not depend on the Bianchi type and can be written with help of (\ref{9}) and (\ref{10}).
\\
The expressions of all the sufficient conditions as function of $G$ and $\omega$ such that the Kretchmann scalar be bounded depends on the Bianchi model. For the Bianchi type $I$ and $V$ model, it is sufficient that the first and second derivatives of $\alpha$, $\beta$ and $\gamma$ be bounded. For the Bianchi type $II$ model, we have an additional conditions, i.e. $\alpha$ have to be bounded. Idem for the Bianchi type $VI_0$ model for which $\alpha$ and $\beta$ have not to diverge. These conditions always imply that $V'V^{-1}$ is bounded. Of course, requiring that the derivatives of $\alpha$, $\beta$ and $\gamma$ do not diverge for the Kretchmann scalar is also sufficient such that the two previous curvature invariants be bounded. Then, the sufficient conditions we found above as function of $G$ and $\omega$ are contained in the requirement that these derivatives be bounded. However, they have been found independently of any Bianchi model. It is why we have considered that it was interesting to deduce them separately.
\subsection{Expression of the sufficient conditions depending on the Bianchi model as function of the scalar field} \label{s32}
In what follows, we examine the previous conditions whose expressions as function of the scalar field depends on the Bianchi model, i.e. $V=e^{\alpha+\beta+\gamma}\not=0$ and the conditions related to the Kretchmann scalar.
\subsubsection{Sufficient conditions such that the  first and second derivatives of $\alpha$, $\beta$ and $\gamma$ be bounded for the Bianchi type $I$ model and the metric functions be non vanishing.} \label{ss31}
The structure constants of the Bianchi type $I$ model are all vanishing. The spatial components of the field equations are:
\begin{eqnarray} 
 &\alpha''=-\alpha' G(G^{-1})'-\frac{1}{2}G(G^{-1})''& \nonumber\\
 &\beta''=-\beta' G(G^{-1})' -\frac{1}{2}G(G^{-1})''& \label{11} \\
 &\gamma''=-\gamma' G(G^{-1})'-\frac{1}{2}G(G^{-1})''& \nonumber
 \end{eqnarray}
We multiply each of them by $G^{-1}$. After an integration, we get:
\begin{equation} \label{12}
\alpha'=(K-1/2(G^{-1})')G
\end{equation}
From a second integration, we deduce:
\begin{equation} \label{12a}
\alpha=K\int G\phi'^{-1}d\phi-1/2\ln G^{-1}
\end{equation}
$K$ is an integration constant. Equivalent expressions can be found for $\beta$ and $\gamma$. With simple considerations, we get some sufficient conditions such that the derivatives of $\alpha$, $\beta$ and $\gamma$ be finite and $V\not = 0$:
\begin{itemize}
\item $K\int G\phi'^{-1}d\phi$ does not diverge toward $-\infty$
\item $G$ is bounded and non vanishing
\item $G(G^{-1})'$ is bounded
\item $G(G^{-1})''$ is bounded
\end{itemize}
Each of them can then be written as function of the scalar field by using equations (\ref{9}) and (\ref{10}). We will study their physical meaning in the last section. In \cite{MimWan95} where a GST with a perfect fluid in the Bianchi type $I$ model is considered, similar conditions for the absence of singularity was found: it has been shown that singularity occurs when $V\rightarrow 0$ and $\phi\rightarrow \infty$.
\subsubsection{Sufficient conditions such that $e^\alpha$ and the first and second derivatives of $\alpha$, $\beta$ and $\gamma$ be bounded for the Bianchi type $II$ model and the metric functions be non vanishing.} \label{ss32}
The non vanishing structure constants are $C^1_{23}=-C^1_{32}=1$. The spatial components of the field equations are written:
\begin{eqnarray}
 &\alpha''=-\alpha' G(G^{-1})'-\frac{1}{2}G(G^{-1})''-1/2e^{4\alpha}& \nonumber \\
 &\beta''=-\beta' G(G^{-1})'-\frac{1}{2}G(G^{-1})''+1/2e^{4\alpha}& \label{14}\\
 &\gamma''=-\gamma' G(G^{-1})'-\frac{1}{2}G(G^{-1})''+1/2e^{4\alpha}& \nonumber
 \end{eqnarray}
In the Einstein frame where the metric functions are related to these of the Brans-Dicke frame by $g_{\mu\nu}=G\tilde{g}_{\mu\nu}$, the solutions of the field equations are well known. They are written $\tilde{\alpha}=1/2\ln(k\cosh^{-1}\left[\tilde{\tau}-\tilde{\tau}_0\right])$,  $\tilde{\beta}=B_0+B_1\tilde{\tau}-1/2\ln(k\cosh^{-1}\left[\tilde{\tau}-\tilde{\tau}_0\right])$ and a similar expression for $\gamma$. $k$, $B_0$, $B_1$ and $\tilde{\tau}_0$ are integration constants. From the Klein-Gordon equation in the Einstein frame, we get $\tilde{\tau}-\tilde{\tau}_0=\int G/\phi'd\phi$. Hence, we deduce the expression of the metric functions in the Brans-Dicke frame:
\begin{eqnarray*}
&\alpha=1/2\ln\{kG\cosh^{-1}(k \int G/\phi'd\phi)\}&\\
&\beta=B_0+B_1\int G\phi'^{-1}d\phi-1/2\ln\{kG^{-1}\cosh^{-1}(k \int G\phi'^{-1}d\phi)&\\
&\gamma=C_0+C_1\int G\phi'^{-1}d\phi-1/2\ln\{kG^{-1}\cosh^{-1}(k \int G\phi'^{-1}d\phi)&\\
\end{eqnarray*}
Thus, some sufficient conditions such that $e^\alpha$ and the first and second derivatives of $\alpha$, $\beta$ and $\gamma$ be bounded for the Bianchi type $II$ model with $V\not = 0$ will be:
\begin{itemize}
\item $\int G\phi'^{-1}d\phi$ is bounded.
\item $G$ is bounded and non vanishing.
\item $G(G^{-1})'$ is bounded.
\item $G(G^{-1})''$ is bounded.
\end{itemize}
These conditions are the same as these of the Bianchi type $I$ model but now $\int G\phi'^{-1}d\phi$ have to be bounded such that $e^\alpha$ stays finite.
\subsubsection{Sufficient conditions such that $e^\alpha$, $e^\beta$, the first and second derivatives of $\alpha$, $\beta$ and $\gamma$ be bounded for the Bianchi type $VI_0$ model and the metric functions be non vanishing.} \label{ss33}
The non vanishing structure constants are $C^1_{23}=-C^1_{32}=C^2_{13}=-C^2_{31}=1$. We will consider the LRS case for which $\alpha=\beta$. The spatial components of the field equations are written:
\begin{eqnarray}
 &\alpha''=-\alpha' G(G^{-1})'-\frac{1}{2}G(G^{-1})''& \label{15}\\
 &\gamma''=-\gamma' G(G^{-1})'-\frac{1}{2}G(G^{-1})''+2e^{4\alpha}& \nonumber
 \end{eqnarray}
The first equation is the same as for the Bianchi type I model. Its solution is then:
\begin{eqnarray*}
&\alpha=K\int G/\phi' d\phi-1/2\ln G^{-1}
\end{eqnarray*}
Putting it in the second equation of (\ref{15}), we get:
\begin{eqnarray*}
&\gamma=-1/2 \ln G^{-1}+(\int G/\phi'd\phi)^2 +2\int e^{4K\int G/\phi'd\phi}/\phi'd\phi &\\
\end{eqnarray*}
Some sufficient conditions such that $e^\alpha$, $e^\gamma$, the first and second derivatives of $\alpha$ and $\gamma$ be bounded with $V\not = 0$ are thus:
\begin{itemize}
\item $\int G\phi'^{-1}d\phi$ is bounded.
\item $G$ is bounded and non vanishing.
\item $G(G^{-1})'$ is bounded.
\item $G(G^{-1})''$ is bounded.
\item $\int e^{4K\int G\phi'^{-1}d\phi}\phi'^{-1}d\phi$ does not tend toward $-\infty$.
\end{itemize}
These conditions are the same as these of the Bianchi type $II$ model except the last one which seems to be specific to the Bianchi type $VI_0$ model.
\subsubsection{Sufficient conditions such that the first and second derivatives of $\alpha$, $\beta$ and $\gamma$ be bounded for the Bianchi type $V$ model and the metric functions be non vanishing.} \label{ss34}
The non vanishing structure constants of this model are $C^2_{21}=-C^2_{12}=C^3_{31}=-C^3_{13}=1$. The spatial components of the field equations are written:
\begin{eqnarray}
 &\alpha''=-\alpha' G(G^{-1})'-\frac{1}{2}G(G^{-1})''+2e^{2\beta+2\gamma}& \nonumber \\
 &\beta''=-\beta' G(G^{-1})'-\frac{1}{2}G(G^{-1})''+2e^{2\beta+2\gamma}& \label{14a}\\
 &\gamma''=-\gamma' G(G^{-1})'-\frac{1}{2}G(G^{-1})''+2e^{2\beta+2\gamma}& \nonumber
 \end{eqnarray}
In the Einstein frame, these three equations are turned into General Relativity equations for the Bianchi type $V$ model whose solutions in the $\tilde{T}$ time defined by $d\tilde{\tau}=d\tilde{T}e^{-\tilde{\beta}-\tilde{\gamma}}=Gd\tau$ have been found by Joseph \cite{Jos66}:
\begin{eqnarray*} 
&e^{2\tilde{\alpha}}=K^2\sinh (2\tilde{T})&\\
&e^{2\tilde{\beta}}=K^2\sinh (2\tilde{T}) \tanh (\tilde{T})^{\sqrt{3}}&\\
&e^{2\tilde{\gamma}}=K^2\sinh(2\tilde{T}) \tanh (\tilde{T})^{-\sqrt{3}}&\\
\end{eqnarray*}
We then calculate that:
\begin{equation}
\tilde{\tau}-\tilde{\tau}_0=1/2K^{-2}\ln \tanh\tilde{T}=\int G\phi'^{-1}d\phi
\end{equation}
The central member of this last expression is defined for $\tilde{T}\in\left[0,+\infty\right[$ and vary from $-\infty$ to $0$. We deduce that the integral diverges when $\tilde{T}\rightarrow 0$ and vanishes when $\tilde{T}\rightarrow +\infty$. So, some sufficient conditions such that the first and second derivatives of $\alpha$, $\beta$ and $\gamma$ be bounded with $V\not = 0$ are:
\begin{itemize}
\item $\tilde{T}$ is bounded and non vanishing, i.e. $\int G\phi'^{-1}d\phi$ do not diverge toward $-\infty$ and is non vanishing.
\item $G$ is bounded and non vanishing.
\item $G(G^{-1})'$ is bounded.
\item $G(G^{-1})''$ is bounded.
\end{itemize}
These conditions are similar to these of the Bianchi type $I$ model but the integral of $G$ with respect to $\tau$ shall be non vanishing. This is in agreement with \cite{MimWan95} where it was noticed that the behaviour of the Bianchi type $V$ model near the singularity is a subset of the Bianchi type $I$ model.\\
\\
A summarise of these results is presented on tables \ref{tab1} and \ref{tab2}.
\section{Applications} \label{s4}
In this section, we use the sufficient conditions of tables \ref{tab1} and \ref{tab2} to find some forms of the functions $G$ and $\omega$ such that the scalar curvature, the Ricci and Kretchmann scalars do not diverge for GST and string inspired theories.
\subsection{Generalised scalar tensor theory} \label{ss41}
The GST is defined by $G^{-1}=\phi$. Lots of papers are devoted to its study \cite{MimWan94,KolEar95,BarPar97, MimWan95}. For this class of theories, we have calculated that:
\begin{eqnarray*}
&G=1/\phi&\\
&G(G^{-1})'\propto \phi^{-1}(3+2\omega)^{-1/2}&\\
&G(G^{-1})''\propto -\omega_\phi\phi^{-1}(3+2\omega)^{-2}&\\
&G\omega\phi'^2\phi^{-1}\propto \omega\phi^{-2}(3+2\omega)^{-1}&\\
\end{eqnarray*} 
From table \ref{tab2} we see that whatever the Bianchi type, $G$ have to be bounded and non vanishing. Hence, $\phi$ is strictly positive or negative and bounded. Since $G(G^{-1})'$ have not to diverge, we shall ask also that $3+2\omega$ be non vanishing. This last function have to be positive such that the energy density of the scalar field be positive in the Einstein frame. Last, $G(G^{-1})''$ have also to be bounded: from what we write for $\phi$, we deduce that it will be verified if it is also the case of $\omega_\phi\omega^{-2}$. All these conditions imply that $G\omega\phi'^2\phi^{-1}$ will stay bounded.\\
A function $\omega$ corresponding to these requirements will be, as instance, $3+2\omega=m+\left[-(i+\phi)(j+\phi)\right]^n$ with $m>0$, $(i,j)<(0,0)$ and $n>1$. The scalar field is then defined on the closed interval $\left[-i,-j\right]$. As $\phi' \propto 1/\sqrt{3+2\omega}$ and $G$ does not diverge, we can show that the integrals of the table \ref{tab2} stay finite. Numerically, we have checked that $\int G\phi'^{-1}d\phi$ is non vanishing. Hence, for all the Bianchi models we have studied, none of the three  curvature invariants diverges.\\
\subsection{String inspired theory} \label{ss42}
The low energy action of the string theory without antisymetric strength field is a HST with $G^{-1}=\omega=e^{-\phi}$. In this application, we will choose $G^{-1}=e^{-\phi}$ and $\omega=e^{-\phi}\Omega(\phi)$. At early time, the compactification of extra dimensions could give birth to physical phenomenon which would be described by such theories \cite{Muk97, Lid96}. It is then interesting to find these which are non singular. We have:
\begin{eqnarray*}
&G=e^{\phi}&\\
&G(G^{-1})'\propto -2\phi_0e^{\phi}(3+2\phi^{-1}\Omega)^{-1/2}&\\
&G(G^{-1})''\propto 4\phi_0^2e^{\phi}(-\Omega+\phi\Omega_{\phi})(3\phi+2\Omega)^{-2}&\\
&G\omega\phi'^2\phi^{-1}\propto e^{2\phi}(3\phi\Omega^{-1}+2)^{-1}&\\
\end{eqnarray*}
$G$ will be bounded and non vanishing if $\phi$ is bounded. From this, we deduce that $G(G^{-1})'$ is bounded and real if $\Omega\phi^{-1}>-3/2$. Then, $G(G^{-1})''$ is bounded if $\phi^2\Omega_\phi\Omega^{-2}$ is finite. All these conditions imply that $G\omega\phi'^2\phi^{-1}$ is always finite. A function $\Omega$ corresponding to these requirements, have the same form as in the previous subsection with $(i,j)<(0,0)$, $n>1$, $m>-3/2$ and $mi^{-1}<3/2$. Then, the same remarks as in subsection \ref{ss41} apply here.
\section{Final remark and conclusion} \label{s5}
In this work, we have determined sufficient conditions such that the scalar curvature, the Ricci and Kretchmann scalars do not diverge for the HST in Bianchi models. It is necessary such that the theory be non singular. These conditions are summarised in table \ref{tab1} and \ref{tab2}.\\
Of course, other types of sufficient conditions can be chosen from this work. As instance, we can replace the two conditions "$G(G^{-1})'$ is bounded" and "$V\not=0$" by "$G(G^{-1})'V^{-1}=G\dot{G}^{-1}$ is bounded". This new condition can be written as a function of $\phi$ by help of (\ref{9}) and the expressions of $V(\phi)$ for each Bianchi model. It is even possible to write the three curvature invariants as functions of $\phi$, $G$, $\omega$ and their derivatives with respect to $\phi$ and to search conditions such that the invariants be bounded directly from these expressions. However, it is not an easy task, particularly when we consider the Kretchmann scalar.\\
What are the physical interpretation of the conditions we have chosen? Whatever the Bianchi models, the gravitation function $G$ have to be bounded and non vanishing. It follows that $(G^{-1})'$ and $(G^{-1})''$ are bounded. Another conditions is that $\int G\phi'^{-1}d\phi=\int G d\tau$ is bounded. Hence, we deduce that the sign of $G$ will not change during time evolution: the gravitation is either attractive or repulsive but can not change its nature. If $G$ tends toward a constant  as it seems to be the case for our present epoch or is asymptotically monotone, as $\int Gd\tau$ is bounded,  $\tau$ is bounded. As $dt=Vd\tau$, it means that if the 3-volume of the Universe diverge or tends asymptotically toward a constant, $t$ will behave in the same way. Hence, a finite asymptotic value of the Universe 3-volume means a finite interval of proper time and an infinite asymptotical value, an open interval of proper time for the Universe. We have also chosen that $G\omega\phi'^2\phi^{-1}$ be bounded. As $G$ do not diverge, if we impose that the solar system tests be respected, i.e. $\omega\rightarrow \infty$, we need then $\phi'\phi^{-1}\rightarrow 0$. As instance, It could be realised if the scalar field tended toward a constant which is a realistic assumption for late times period. Note that the condition on $G\omega\phi'^2\phi^{-1}$ is not independent of the others: it is a consequence of the Klein-Gordon equation, the fact that $G^{-1}$ is non vanishing, bounded and $(G^{-1})'$ is bounded. This fact has been observed in the applications of section \ref{s4}.\\
Finally, the conditions we have established such that the three curvature invariants we studied be finite can be summarised in four simple points for the Bianchi types I, II, V and VI models:
\begin{itemize}
\item $G$ is bounded and non vanishing
\item $(G^{-1})'$ and $(G^{-1})''$ which can be expressed with equations (\ref{9}) and (\ref{10}) are bounded
\item $\int Gd\tau$ does not diverge toward $-\infty$ or/and $+\infty$ depending on the Bianchi models
\item For the Bianchi type $VI_0$ model there is a special conditions, i.e. $\int e^{4K\int Gd\tau}d\tau$ does not diverge.
\end{itemize}
From these conditions and the fields equations obtained for each Bianchi model, we deduce that the $n^{th}$ derivatives of each metric function with respect to $\tau$ will be bounded if the $n^{th}$ derivative of $G$ with respect to $\tau$ is bounded. This derivative can be calculated as a function of $G$, $\omega$ and their derivatives with respect to the scalar field with help of the recursive relation $d^{n}G/d\tau^{n}=d\left[d^{n-1}G/d\tau^{n-1}\right]/d\phi \phi'$ and the relation (\ref{5}). Adding this last conditions to the four previous one, this ensures then that each curvature invariant is bounded.\\
Each of these conditions can be expressed with $G$, $\omega$ and their derivatives with respect to the scalar field. Hence we have achieved the goal we fixed at the beginning of the paper: we have found a simple set of sufficient conditions such that the invariant curvatures of the HST be bounded. The theories defined by these conditions could then be interesting theories to represent asymptotical behaviour of an anisotropic Universe if we assume that the singularity must be avoided.\\

A similar work has been carried out by Rama \cite{Ram97}, with the GST with a perfect fluid for the FLRW models. In this last paper, as sufficient conditions such that none of the curvature invariant diverges, it was chosen that the invert of the scale factor, $e^A$, and the successive derivatives of $A$ with respect to the proper time $t$ be bounded. If we exclude the conditions specific to the presence of matter, the others were written:
\begin{itemize}
\item $e^{-A}$ is bounded
\item $\dot{\phi}\phi^{-1}$ is bounded
\item $\omega\dot{\phi}^2\phi^{2}$ is bounded
\item $\phi^n(3+2\omega)^{-1}d^n(3+2\omega)/d\phi^n(\dot{\phi}\phi^{-1})^n$ is bounded
\end{itemize}
We have recovered the first one by writing that the 3-volume be non vanishing. It implies also that $\int Gd\tau$ is bounded. Since we have chosen that $G(G^{-1})'$ and $V^{-1}$ are bounded, it means that there product is bounded too. This implies the second condition since $dt=Vd\tau$ and $\phi$ should be replaced by $G^{-1}$ for the hyperextended theory. However the reverse is false. In this way, the sufficient conditions we have chosen are more restrictive for the functions $G$ and $\omega$ than these of Rama. We can recover the third condition of Rama in the same way since we have assumed that $V^{-1}$ and $G\omega\phi'^2\phi^{-1}$ were bounded. the fourth condition come from the fact that the successive derivatives of $A$ have to be bounded and should be related with the condition on the finiteness of $d^{n}G/d\tau^{n}$. Hence, by matching the conditions chosen in this work, we can recover all the conditions of \cite{Ram97} which are not concerned by the presence of matter. The main difference comes from the fact that we have replaced $\phi$ by $G^{-1}$, i.e it arises because we have considered an hyperextended rather than a generalised scalar tensor theory. It is a difference of physical order. There is also another difference which is of geometrical order. The only condition present in this work and not in \cite{Ram97} is the one for the Bianchi type $VI_0$ model, implying that $\int e^{4K\int G\phi'^{-1}d\phi}\phi'^{-1}d\phi$ is bounded. It seems to be a specific condition characterising this model since it has no equivalent in the other Bianchi models. It would explain why it does not appear in the paper of Rama since Bianchi type $VI_0$ model can not be related to a FLRW one. Hence, this paper complete \cite{Ram97} by extended some of its results in the HST for Bianchi models which was one of the issues evoked in its conclusion.

\begin{table}[p]
\begin{indented}
\item[]\begin{tabular}{@{}ll}
\br
curvature invariant & Bounded quantities \\
\mr
$R$			    & $G(G^{-1})''$, $G\omega\phi'^2\phi^{-1}$, $\alpha'$, $\beta'$, $\gamma'$\\
$R_{\mu\nu}R^{\mu\nu}$ & $G(G^{-1})''$, $G\omega\phi'^2\phi^{-1}$, $G(G^{-1})'$, $\alpha'$, $\beta'$, $\gamma'$\\
\mr
 $R_{\alpha\beta\mu\nu}R^{\alpha\beta\mu\nu}$& \\
 $I$ et $V$	& $\alpha'$, $\beta'$, $\gamma'$, $\alpha''$, $\beta''$, $\gamma''$\\
 $II$ 		& $\alpha'$, $\beta'$, $\gamma'$, $\alpha''$, $\beta''$, $\gamma''$, $e^\alpha$\\
 $VI_0$	& $\alpha'$, $\beta'$, $\gamma'$, $\alpha''$, $\beta''$, $\gamma''$, $e^\alpha$,$e^\gamma$\\
\br
\end{tabular}
\end{indented}
\caption{When $V\not =0$, it is sufficient that these quantities be bounded such that the curvature invariants do not diverge. For the scalar curvature and the Ricci scalar, these conditions are independent on the considered Bianchi models.}
\label{tab1}
\end{table}

\begin{table}[p]
\begin{indented}
\item[]\begin{tabular}{@{}ll}
\br
Model 		& Conditions  \\
\mr
$I$			& $G$, $G(G^{-1})'$, $G(G^{-1})''$ are bounded, $G$ is non vanishing\\
			& $K\int G\phi'^{-1}d\phi$ does not diverge toward $-\infty$\\
 $II$			& $G$, $G(G^{-1})'$, $G(G^{-1})''$ is bounded, $G$ is non vanishing\\
			& $\int G\phi'^{-1}d\phi$ is bounded\\
$VI_0$ 		& $G$, $G(G^{-1})'$, $G(G^{-1})''$ is bounded, $G$ is non vanishing\\
LRS			& $\int G\phi'^{-1}d\phi$ is bounded\\
			& $\int e^{4K\int G\phi'^{-1}d\phi}\phi'^{-1}d\phi$  does not tend toward $-\infty$\\
$V$			& $G$, $G(G^{-1})'$, $G(G^{-1})''$ is bounded, $G$ is non vanishing\\
			&$\int G\phi'^{-1}d\phi$ is non vanishing and does not diverge toward $-\infty$\\
\br
\end{tabular}
\end{indented}
\caption{Sufficient conditions such that the Kretchmann scalar be bounded and the 3-volume be non vanishing for the Bianchi type $I$, $II$, $VI_0$ and $V$ models.}
\label{tab2}
\end{table}

\end{document}